\documentclass{article}

\usepackage[preprint]{neurips_2026}

\usepackage[utf8]{inputenc} 
\usepackage[T1]{fontenc}    
\usepackage{hyperref}       
\usepackage{url}            
\usepackage{booktabs}       
\usepackage{amsfonts}       
\usepackage{nicefrac}       
\usepackage{microtype}      
\usepackage[table]{xcolor}  
\usepackage{svg}

\usepackage{multirow}
\usepackage[table]{xcolor}
\usepackage{pifont}
\usepackage{makecell}

\usepackage{graphicx}
\usepackage{amsmath}

\usepackage{pgfplots}
\pgfplotsset{compat=1.18}
\usepackage[most]{tcolorbox}

\title{SIREM: Speech-Informed MRI Reconstruction with Learned Sampling}

\author{%
\textbf{Md Hasan}$^{1}$\thanks{Corresponding authors: \texttt{md.hasan@fau.de}; \texttt{paula.andrea.perez@fau.de}.} \quad
\textbf{Nyvenn Castro}$^{2}$ \quad
\textbf{Daiqi Liu}$^{1}$ \quad
\textbf{Lukas Mulzer}$^{1}$ \quad
\textbf{Jana Hutter}$^{3}$ \\ \quad 
\textbf{Jonghye Woo}$^{4}$ \quad
\textbf{Moritz Zaiss}$^{2}$ \quad
\textbf{Andreas Maier}$^{1}$ \quad
\textbf{Paula A. P\'erez-Toro}$^{1*}$ \\[7pt]
\footnotesize
$^{1}$Pattern Recognition Lab, Friedrich-Alexander-Universit\"at Erlangen-N\"urnberg, \footnotesize Erlangen, Germany \\[-1pt]
\footnotesize
$^{2}$Institute of Radiology, University Hospital Erlangen, Friedrich-Alexander-Universität Erlangen-Nürnberg,\\\footnotesize Erlangen, Germany \\[-1pt]
\footnotesize
$^{3}$Institut f\"ur Informationsverarbeitung, Leibniz Universit\"at Hannover, Hannover, Germany \\[-1pt]
\footnotesize
$^{4}$Department of Radiology, Harvard Medical School and Massachusetts General Hospital, Boston, MA, USA
}

\begin{document}

\maketitle

\begin{abstract}

Real-time magnetic resonance imaging (rtMRI) of speech production enables non-invasive visualization of dynamic vocal-tract motion and is valuable for speech science and clinical assessment. However, rtMRI is fundamentally constrained by trade-offs among spatial resolution, temporal resolution, and acquisition speed, often leading to undersampled k-space measurements and degraded reconstructions. We propose SIREM, a speech-informed MRI reconstruction framework that uses synchronized speech as a cross-modal prior. The central idea is that vocal-tract configurations during speech are correlated with the produced acoustics, making part of the image content predictable from audio. SIREM models each frame as a fusion of an audio-driven component and an MRI-driven component through a spatial weighting map. The audio branch predicts articulator-related structure from speech, while the MRI branch reconstructs complementary content from measured k-space data. We further introduce a learnable soft weighting profile over spiral arms, enabling a differentiable study of how k-space arm usage interacts with speech-informed fusion. This yields a unified multimodal formulation that combines audio-driven prediction, MRI reconstruction, and sampling adaptation. We evaluate SIREM on the USC speech rtMRI benchmark against standard baselines, including gridding, wavelet-based compressed sensing, and total variation. SIREM introduces a speech-informed reconstruction paradigm that operates in a substantially higher-throughput regime than iterative methods while preserving anatomically plausible vocal-tract structure. These results establish an initial benchmark for multimodal speech-informed rtMRI reconstruction and highlight the potential of synchronized speech as an auxiliary prior for fast reconstruction. The source code is available at \url{https://github.com/mdhasanai/SIREM}

\end{abstract}

\section{Introduction}
Real-time magnetic resonance imaging (rtMRI) visualizes the dynamics of speech production by capturing the full midsagittal vocal tract, tongue, lips, velum, pharynx, and laryngeal structures without ionizing radiation or articulatory constraint. Unlike electromagnetic articulography~\cite{berry2011accuracy} or ultrasound~\cite{klein2013multidimensional}, this makes rtMRI uniquely suited for articulatory and phonetic analysis~\cite{ramanarayanan2018analysis}, as well as clinical assessment of conditions such as velopharyngeal insufficiency, dysarthria, and post-surgical speech impairments~\cite{hagedorn2017characterizing}.

To satisfy temporal-resolution constraints, speech rtMRI reconstructs only a fraction of k-space per frame, turning reconstruction into a strongly ill-posed inverse problem~\cite{le_online_2026}. Standard parallel imaging reconstruction methods such as SENSE~\cite{pruessmann1999sense} and GRAPPA~\cite{griswold2002generalized} exploit coil and k-space redundancies. 
Speech-specific dynamic reconstruction methods further exploit temporal structure through low-rank, manifold, or subspace models~\cite{fu2015high,rusho2024prospectively,cao2025self,le_online_2026}. These approaches have been effective, but they rely primarily on measurement-domain, image-domain, or temporal priors and do not explicitly use the synchronized speech waveform available during acquisition.

Speech rtMRI is, however, intrinsically multimodal: each acquisition includes a synchronized audio waveform in addition to the MRI measurements~\cite{lim2021multispeaker}. The acoustic signal is not merely an annotation, but a physical consequence of the vocal-tract configuration that the MRI seeks to image. Articulatory studies show that the tongue, lips, jaw, and velum shape the spectral envelope, formant trajectories, and voicing structure of speech~\cite{ramanarayanan2018analysis}, and recent audio-conditioned rtMRI synthesis and inversion studies suggest that substantial articulatory information can be recovered from speech alone~\cite{udupa2023real,nguyen2025speech2rtmri,shi2025speech,perez2026speech}. Thus, the audio channel contains structured cues about the underlying image content that are typically discarded by reconstruction pipelines.

These observations motivate us to formulate speech rtMRI reconstruction as a multimodal estimation problem in which synchronized speech provides an auxiliary prior on anatomically relevant regions. Hence, we propose SIREM, a hybrid framework that fuses an audio-driven image estimate with an MRI-driven reconstruction via a spatial weighting map, and incorporates a learnable soft spiral-arm reweighting profile to study k-space arm usage under speech-informed fusion.

Our contributions are:

\begin{itemize}
\item We formulate speech rtMRI reconstruction as a multimodal inverse problem in which synchronized audio serves as a cross-modal prior on vocal-tract structure.
\item We propose SIREM, a hybrid framework that fuses audio-driven and MRI-driven estimates via a spatial weighting map and a learnable soft spiral-arm reweighting profile.
\item We establish an initial benchmark for multimodal speech-informed rtMRI reconstruction on the USC corpus, comparing against gridding, wavelet, and total variation baselines.

\end{itemize}

SIREM does not uniformly outperform classical iterative reconstruction:
wavelet retains the strongest $\ell_2$-style distortion scores in every evaluation protocol. Yet demonstrates that multimodal direct reconstruction is feasible, that synchronized audio is a useful auxiliary prior, and that the resulting throughput regime is qualitatively distinct from iterative approaches.

\section{Related Work}
Speech rtMRI has become a central modality for observing the moving vocal tract because it captures the full midsagittal anatomy, including structures that are difficult to measure with electromagnetic articulography or ultrasound. Public speech-rtMRI resources, particularly the USC 75-speaker corpus, provide synchronized audio, raw multi-coil non-Cartesian k-space, and reconstructed image sequences~\citep{lim2021multispeaker}. In this work, we use the 75-Speaker Annot-16 subset, which provides manual articulatory annotations and supplies the spatial weighting masks used in the reported experiments~\citep{shi202575}.

Classical accelerated MRI reconstruction is built on parallel imaging and compressed sensing. SENSE~\cite{pruessmann1999sense} and GRAPPA~\cite{griswold2002generalized} exploit spatial encoding redundancy across receiver coils, while compressed-sensing MRI combines data consistency with sparsity-promoting priors, such as wavelets and total variation~\citep{lustig2007sparse}. Speech-specific rtMRI reconstruction has also explored dynamic priors including temporal finite-difference regularization, low-rank-plus-sparse models, manifold regularization, and subspace-based reconstruction for vocal-tract imaging~\citep{fu2015high, rusho2024prospectively, cao2025self}. These approaches improve reconstruction by exploiting measurement-domain or temporal structure, but they do not use the synchronized waveform as an explicit prior.

More recent MRI reconstruction has shifted toward physics-informed deep learning and unrolled optimization in accelerated MRI~\citep{aggarwal2018modl, hammernik2018variational,sriram2020end}. A related direction studies joint optimization of sampling and reconstruction, treating acquisition design as a learnable component rather than a fixed heuristic~\citep{bahadir2019learning,zhang2019reducing}. SIREM is conceptually related to this line of work, but addresses a distinct setting in which synchronized audio provides side information for reconstruction from raw spiral k-space.

Synchronized speech has also been used for vocal-tract modeling in generation, inversion, and segmentation tasks, audio-conditioned rtMRI synthesis, diffusion-based speech-to-rtMRI generation, and audio-assisted vocal-tract segmentation~\citep{udupa2023real,nguyen2025speech2rtmri,perez2026speech,shi2025speech,jain2024multimodal,liu2025vocsegmri}. These studies support the premise that speech contains substantial articulatory information, yet the inclusion in speech imaging reconstruction pipelines remains largely unexplored. SIREM instead targets the complementary problem of using synchronized speech as an auxiliary prior for reconstructing undersampled speech rtMRI directly from raw k-space measurements.

\section{SIREM}

We consider speech-informed reconstruction for real-time MRI from undersampled non-Cartesian k-space measurements. Our central hypothesis is that part of the articulatory configuration is predictable from synchronized speech audio, while the remaining anatomy must be recovered from measured MRI data. Following this decomposition, we model each frame as a fusion of an audio-driven estimate and an MRI-driven reconstruction, and we learn a soft sampling policy over spiral arms to study whether acoustically predictable regions permit more aggressive undersampling.

\begin{figure}[!htpb]
    \centering
    \includegraphics[width=0.95\linewidth]{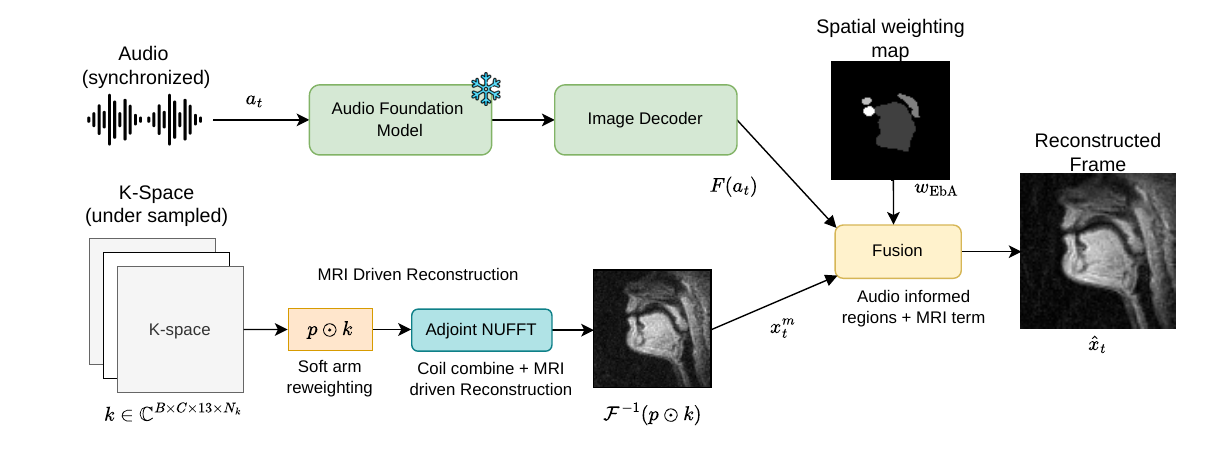}
\caption{Overview of SIREM. The audio branch maps synchronized speech $a_t$ to an audio-driven estimate $x_t^{a}$, while the measurement branch applies the soft spiral-arm profile $p$ to undersampled multi-coil k-space $k_t$ and reconstructs the MRI-driven estimate $x_t^{m}$. Pixelwise fusion under the explained-by-audio map $w_{\mathrm{EbA}}$ yields the final frame $\hat{x}_t$.
}
\label{fig:sirem_architecture}

\end{figure}

Figure~\ref{fig:sirem_architecture} illustrates the resulting computational graph. The audio prior path maps the synchronized speech segment $a_t$ to an audio-driven estimate $x_t^{a}$ through a frozen HuBERT encoder and a trainable image decoder. In parallel, the measurement branch applies the learnable soft spiral-arm profile $p$ to the acquired multi-coil k-space $k_t$ and reconstructs the MRI-driven estimate $x_t^{m}=\mathcal{F}^{-1}(p \odot k_t)$ by adjoint NUFFT reconstruction and SENSE coil combination. The final frame $\hat{x}_t$ is then obtained by pixelwise fusion of $x_t^{a}$ and $x_t^{m}$ under the explained-by-audio map $w_{\mathrm{EbA}}$.

\subsection{Problem formulation}

Let $x_t \in [0,1]^{H \times W}$ denote the target midsagittal frame, $a_t \in \mathbb{R}^{L}$ the synchronized speech segment, and $k_t \in \mathbb{C}^{C \times R \times N_k}$ the acquired multi-coil spiral k-space for frame $t$, where $C$ is the number of coils, $R=13$ is the number of spiral arms per full rotation, and $N_k$ is the number of readout samples per arm. During training, these frame-wise quantities are batched over samples. The goal is to estimate $x_t$ from the paired observation $(k_t,a_t)$.

SIREM is formulated as three coupled operators: an audio prediction branch, an MRI reconstruction branch, and a spatial fusion map. The audio branch maps synchronized speech to an audio-driven estimate,
\begin{equation}
x_t^{a} = F(a_t),
\label{eq:audio_branch}
\end{equation}
where $F(\cdot)$ denotes the audio foundation model followed by the image decoder. The MRI branch applies a learnable soft spiral-arm profile $p \in [0,1]^R$ to the acquired measurements and reconstructs an MRI-driven estimate,
\begin{equation}
x_t^{m} = \mathcal{F}^{-1}(p \odot k_t),
\label{eq:mri_branch}
\end{equation}
where $\odot$ denotes elementwise multiplication with broadcasting over coils and readout samples, and $\mathcal{F}^{-1}$ denotes the the SENSE-weighted adjoint-NUFFT-based reconstruction and coil-combination operator.

The final reconstruction is obtained by pixelwise fusion of the two estimates,
\begin{equation}
\hat{x}_t
=
w_{\mathrm{EbA}} \odot x_t^{a}
+
\left(1-w_{\mathrm{EbA}}\right)\odot x_t^{m},
\label{eq:fusion}
\end{equation}
where $w_{\mathrm{EbA}} \in [0,1]^{H \times W}$ is the explained-by-audio spatial weighting map. Larger values of $w_{\mathrm{EbA}}$ increase the contribution of the audio-driven estimate in acoustically predictable regions, whereas smaller values preserve measurement-driven recovery through $x_t^{m}$.

\subsection{Dataset}
Experiments are conducted on the USC speech rtMRI corpus~\cite{lim2021multispeaker}, which provides synchronized speech audio, raw k-space measurements, and reconstructed midsagittal vocal-tract image sequences acquired on a 1.5T scanner using a non-Cartesian spiral trajectory. Since the proposed framework requires anatomical masks to define the explained-by-audio prior, we use the USC-16 subset \cite{shi202575}, a curated benchmark derived from USC-75 with manual articulatory annotations. We adopt a subject-independent split with 10 training subjects, 2 validation subjects, and 4 test subjects. The validation set is used for model selection and hyperparameter tuning, whereas the test set is used only for final evaluation.

\paragraph{Preprocessing and temporal alignment.} Each training instance consists of synchronized k-space measurements, trajectories, speech audio, reference reconstructions, and segmentation masks. Audio is resampled to 16~kHz, and image-domain inputs are reoriented to a consistent midsagittal view, resized to $84 \times 84$, and normalized to $[0,1]$.

The acquisition uses 13 spiral arms per full rotation. In our setting, each training sample corresponds to one 13-arm acquisition window, yielding an effective temporal resolution of 12.81 fps. Since the provided reference reconstructions and segmentation masks are available at the 2-arm rate (83.28 fps), we align each 13-arm sample with the temporally centered high-rate reference frame. Let $c_t$ denote the temporal center of the $t$-th 13-arm acquisition window. The aligned reference timestamp is defined as
\begin{equation}
\tau_t = \arg\min_{\tau \in \mathcal{T}_{2\text{-arm}}} |\tau - c_t|
\label{eq:temporal_alignment}
\end{equation}
where $\mathcal{T}_{2\text{-arm}}$ denotes the timestamps of the 2-arm reference sequence. The associated audio segment is extracted around the same center using a symmetric context window,
\begin{equation}
a_t = a[c_t-\delta : c_t+\delta]
\label{eq:audio_window}
\end{equation}

\paragraph{K-space representation and sensitivity estimation.}
For each utterance, the acquired spiral arms are grouped into consecutive blocks of 13, yielding frame-wise k-space tensors of shape $[T,C,13,N]$, where $T$ denotes the number of frames, $C$ the number of coils, and $N$ the number of readout samples per arm. For frame $t$, we write
\begin{equation}
k_t \in \mathbb{C}^{C \times 13 \times N},
\label{eq:kspace_tensor}
\end{equation}
where the second dimension indexes the 13 spiral arms and the third dimension indexes the readout samples along each arm. The corresponding non-Cartesian sampling trajectories are arranged as
\begin{equation}
\omega_t \in \mathbb{R}^{13 \times N \times 2},
\label{eq:trajectory_tensor}
\end{equation}
where each trajectory sample specifies a two-dimensional k-space coordinate. Density compensation weights are taken from the trajectory metadata. K-space values are normalized by their maximum magnitude,
\begin{equation}
\tilde{k}_t = \frac{k_t}{\max_{c,i,n} |k_t[c,i,n]|},
\label{eq:kspace_normalization}
\end{equation}
and trajectories are scaled to the reconstruction grid. Coil sensitivity maps are estimated once per utterance by averaging k-space over full rotations, reconstructing low-resolution coil images with an adjoint NUFFT, and applying Walsh-based sensitivity estimation~\cite{walsh2000adaptive}.

\subsection{Speech-informed reconstruction model}

\paragraph{Audio-driven branch.}

The audio-driven branch implements the mapping $F$ in Eq.~\ref{eq:audio_branch} and produces the estimate $x_t^{a}$ from the synchronized speech segment $a_t$. Concretely, HuBERT extracts contextual speech features, which are temporally pooled and decoded into an image-domain estimate in $\mathbb{R}^{H \times W}$. The resulting $x_t^{a}$ is intended to capture articulator configurations whose geometry is strongly coupled to acoustics, particularly the tongue, lips, and velum.

The audio-driven branch predicts an articulator-focused image estimate from synchronized speech. Let $a_t$ denote the waveform segment centered at time $t$. We first extract contextual speech features using a frozen HuBERT encoder~\cite{hsu2021hubert},
\begin{equation}
h_t = E_{\mathrm{HuBERT}}(a_t), \qquad h_t \in \mathbb{R}^{L \times d},
\label{eq:hubert_features}
\end{equation}
where $L$ is the number of latent time steps and $d$ is the feature dimension. The latent sequence is then mean-pooled over time,
\begin{equation}
\bar{h}_t = \frac{1}{L}\sum_{\ell=1}^{L} h_{t,\ell},
\label{eq:hubert_pooling}
\end{equation}
and mapped to image space by a lightweight multilayer decoder,
\begin{equation}
x_t^{a} = F(a_t) = D_{\theta}(\bar{h}_t), \qquad x_t^{a} \in \mathbb{R}^{H \times W},
\label{eq:audio_branch2}
\end{equation}
with $H=W=84$ in our implementation. The decoder consists of three fully connected layers with hidden dimensions 1024 and 2048, LayerNorm, GELU activations, and dropout, followed by a final linear projection to the image grid. This branch is intended to capture articulator configurations that are strongly coupled to acoustics, particularly the tongue, lips, and velum.

\paragraph{MRI-driven branch.}

The MRI-driven branch produces the estimate $x_t^{m}$ from undersampled multi-coil k-space. Let $p \in (0,1)^R$ denote the learnable soft weighting profile over the $R=13$ spiral arms. For frame $t$, the weights are broadcast across coils and readout samples and applied to the k-space tensor as
\begin{equation}
\tilde{k}_t[c,i,n] = p_i \, k_t[c,i,n],
\label{eq:soft_arm_weighting}
\end{equation}
where $c$ indexes coils, $i$ indexes spiral arms, and $n$ indexes readout samples along each arm. The weighted measurements are then mapped to image space with an adjoint NUFFT-based gridding operator~\cite{ong2019sigpy}. Using coil sensitivity maps $S_c$, the MRI-driven reconstruction is written as
\begin{equation}
x_t^{m}
=
\mathcal{F}^{-1}(p \odot k_t)
=
\sum_{c=1}^{C} S_c^{*}\,\mathcal{F}^{-1}_{\mathrm{NUFFT}}\!\left(\tilde{k}_t^{(c)}\right),
\label{eq:mri_branch2}
\end{equation}
where $\mathcal{F}^{-1}_{\mathrm{NUFFT}}$ denotes the adjoint non-Cartesian Fourier operator and $S_c^{*}$ is the complex conjugate of the sensitivity map for coil $c$. The resulting reconstruction is normalized to $[0,1]$ on a per-sample basis before fusion with the audio-driven estimate. This branch is responsible for preserving image content that remains anchored to measured MRI data.

\paragraph{Spatial weighting map.}

The explained-by-audio map modulates the relative contribution of the audio-driven estimate $x_t^{a}$ and the MRI-driven estimate $x_t^{m}$ at each pixel. Let
\begin{equation}
w_{\mathrm{EbA}} \in [0,1]^{H \times W}
\label{eq:weight_map}
\end{equation}
denote the explained-by-audio map, where larger values indicate regions assumed to be more predictable from synchronized speech. The final reconstruction is obtained by pixelwise fusion,
\begin{equation}
\hat{x}_t
=
w_{\mathrm{EbA}} \odot x_t^{a}
+
(1-w_{\mathrm{EbA}}) \odot x_t^{m},
\label{eq:fusion_final}
\end{equation}
where $\odot$ denotes elementwise multiplication. Larger values of $w_{\mathrm{EbA}}$ increase reliance on the acoustically predictable component, whereas smaller values preserve measurement-driven recovery through $x_t^{m}$. In the reported experiments, $w_{\mathrm{EbA}}$ is not learned jointly but is instead derived from segmentation masks that highlight articulator regions plausibly predictable from acoustics. This imposes an explicit anatomical prior, makes the prior anatomically interpretable, stabilizes training in the low-data USC-16 regime, and isolates the contribution of speech-informed fusion. Although the implementation also supports a learnable predictor for $w_{\mathrm{EbA}}$ from fused audio and MRI features, that variant is not used in the reported experiments.

\subsection{Learned sampling policy}

The learnable spiral-arm profile $p$ is optimized jointly with the reconstruction model and enters the method through the MRI-driven pathway $x_t^{m}=\mathcal{F}^{-1}(p \odot k_t)$. We parameterize this profile with trainable logits $\ell \in \mathbb{R}^{13}$ and define
\begin{equation}
p = \sigma(\ell), \qquad p \in (0,1)^{13},
\label{eq:psoft}
\end{equation}
where $\sigma(\cdot)$ denotes the sigmoid function applied elementwise. The resulting vector assigns a continuous weight to each of the 13 spiral arms and provides a differentiable mechanism for modulating their contribution to reconstruction. In the reported experiments, $p$ is applied retrospectively to already acquired k-space data and should therefore be interpreted as a differentiable surrogate for arm importance rather than as a prospectively deployed acquisition policy. Although the implementation also supports hard top-$k$ arm selection via a straight-through estimator, all reported results use the soft weighting profile throughout training and inference.

\subsection{Training objective}

Let $\mathcal{D}=\{(k_t,a_t,x_t,w_{\mathrm{EbA}})\}_{t=1}^{T}$ denote the training set, where $k_t$ is the multi-coil k-space measurement, $a_t$ is the synchronized audio segment, $x_t$ is the reference frame, and $w_{\mathrm{EbA}}$ is the explained-by-audio spatial weighting map. The objective optimizes the fused reconstruction $\hat{x}_t$ while regularizing both the spiral-arm profile $p$ and the spatial weighting map $w_{\mathrm{EbA}}$. We minimize the expected loss
\begin{equation}
\mathcal{L}
=
\mathbb{E}_{(k_t,a_t,x_t,w_{\mathrm{EbA}})\sim\mathcal{D}}
\left[
\mathcal{L}_{\mathrm{recon}}
+ \alpha \mathcal{L}_{\mathrm{psf}}
+ \beta \mathcal{L}_{\mathrm{budget}}
+ \gamma \mathcal{L}_{\mathrm{mask}}
\right].
\label{eq:total_loss}
\end{equation}
The reconstruction term enforces fidelity between the fused estimate $\hat{x}_t$ and the reference frame $x_t$:
\begin{equation}
\mathcal{L}_{\mathrm{recon}}
=
\|\hat{x}_t - x_t\|_2^2.
\label{eq:recon_loss}
\end{equation}
To regularize the learned spiral-arm weighting profile $p$, we penalize the energy of its inverse Fourier transform,
\begin{equation}
\mathcal{L}_{\mathrm{psf}}
=
\mathrm{mean}\left(\left|\mathcal{F}^{-1}(p)\right|^2\right),
\label{eq:p_reg}
\end{equation}
which discourages undesirable point-spread behavior induced by highly irregular arm weights. To encourage sparse arm usage around a target effective budget $K$, we define
\begin{equation}
\mathcal{L}_{\mathrm{budget}}
=
\left(\sum_{i=1}^{13} p_i - K\right)^2
+
\frac{1}{13}\sum_{i=1}^{13} p_i,
\label{eq:sparsity}
\end{equation}
where $K=2$ in all reported experiments. Finally, to regularize the spatial weighting map, we penalize the Fourier energy of its complement,
\begin{equation}
\mathcal{L}_{\mathrm{mask}}
=
\mathrm{mean}\left(
\left|
\mathcal{F}\!\left(1-w_{\mathrm{EbA}}\right)
\right|^2
\right),
\label{eq:w_reg}
\end{equation}
which favors smoother transitions between audio-dominant and MRI-dominant regions. In all experiments, we set $\alpha=0.1$, $\beta=0.01$, and $\gamma=0.01$.

\paragraph{Implementation details}
The HuBERT encoder is frozen during training, and only the audio decoder and sampling logits are optimized. We use AdamW with learning rate $10^{-4}$, weight decay $10^{-5}$, batch size 8, and cosine annealing over 20 epochs. Gradients are clipped to a maximum norm of 1.0. Validation is performed every 5 epochs, and the final model is selected according to validation peak signal-to-noise ratio (PSNR). All experiments were conducted on a single NVIDIA Quadro RTX 5000 GPU (16 GB).

\section{Experiments and Results}

We evaluate SIREM against three reconstruction baselines commonly used in accelerated MRI: direct gridding, wavelet-based compressed sensing, and total-variation reconstruction. These comparisons test whether synchronized speech can serve as a useful prior under undersampled acquisition. Reconstruction quality is assessed using both distortion-based and perceptual metrics, including peak signal-to-noise ratio (PSNR), structural similarity index measure (SSIM)~\cite{wang2004image}, normalized mean squared error (NMSE), mean squared error (MSE), normalized root mean squared error (NRMSE), high-frequency error norm (HFEN), learned perceptual image patch similarity (LPIPS)~\cite{zhang2018unreasonable}, visual information fidelity (VIF)~\cite{sheikh2006image}, and Fr\'echet inception distance (FID)~\cite{heusel2017gans}. Metrics are computed sequence-wise and averaged over the evaluation set.

\subsection{Results}

We report results under two evaluation protocols. In the first, reconstructions are compared against the reference images distributed with the USC Annot-16 benchmark at the native 13-arms-per-frame acquisition rate (Table~\ref{tab:unified_reconstruction_comparison}).

\begin{table*}[!htpb]
\centering
\caption{
Unified quantitative comparison of reconstruction methods under three evaluation targets:
(1) total-variation (TV) reconstructions provided in the USC-16 benchmark,
(2) fully sampled gridding reconstructions,
and (3) wavelet reconstructions.
Results highlight how reconstruction quality changes depending on the reference target.
}
\setlength{\tabcolsep}{9pt} 
\resizebox{\textwidth}{!}{
\begin{tabular}{lccccccc}
\toprule
\midrule
\textbf{Method} & \textbf{SSIM} $\uparrow$ & \textbf{PSNR} $\uparrow$ & \textbf{HFEN} $\downarrow$ & \textbf{NRMSE} $\downarrow$ & \textbf{LPIPS} $\downarrow$ & \textbf{VIF} $\uparrow$ & \textbf{FID} $\downarrow$ \\
\midrule
\multicolumn{8}{c}{\textbf{Target: Total Variation (TV)}} \\
\midrule
Gridding & 0.731 & 26.76 & 0.336 & 0.046 & 0.071 & 0.877 & 89.36 \\
Wavelet & 0.748 & 26.93 & 0.317 & 0.045 & 0.064 & 0.853 & 75.38 \\
\rowcolor{gray!8}
SIREM ($w/o$ audio) & 0.684 & 24.32 & 0.431 & 0.062 & 0.118 & 0.828 & 123.85 \\
\rowcolor{gray!8}
SIREM & 0.702 & 24.69 & 0.396 & 0.058 & 0.110 & 0.847 & 121.16 \\

\midrule
\multicolumn{8}{c}{\textbf{Target: Gridding}} \\
\midrule
Wavelet & 0.960 & 39.19 & 0.160 & 0.011 & 0.005 & 0.624 & 19.25 \\
TV & 0.847 & 33.77 & 0.333 & 0.020 & 0.054 & 0.294 & 72.65 \\
\rowcolor{gray!8}
SIREM ($w/o$ audio) & 0.833 & 27.75 & 0.289 & 0.046 & 0.039 & 0.867 & 33.80 \\
\rowcolor{gray!8}
SIREM & 0.881 & 28.71 & 0.213 & 0.036 & 0.021 & 0.912 & 27.33 \\

\midrule
\multicolumn{8}{c}{\textbf{Target: Wavelet}} \\
\midrule
Gridding & 0.932 & 33.08 & 
0.155 & 0.022 & 0.007 & 0.965 & 23.84 \\
TV & 0.748 & 26.93 & 0.326 & 0.045 & 0.064 & 0.768 & 75.38 \\
\rowcolor{gray!8}
SIREM ($w/o$ audio) & 0.808 & 27.05 & 0.293 & 0.047 & 0.039 & 0.870 & 41.71 \\
\rowcolor{gray!8}
SIREM & 0.830 & 27.47 & 0.257 & 0.042 & 0.030 & 0.891 & 38.63 \\
\midrule
\bottomrule
\end{tabular}
}
\label{tab:unified_reconstruction_comparison}
\end{table*}

The results reveal a fidelity-throughput trade-off rather than a uniform quality gain. Table~\ref{tab:unified_reconstruction_comparison} reports performance under three reference targets: the USC Annot-16 benchmark reference, fully sampled gridding, and wavelet reconstructions. Across targets, wavelet and gridding remain strongest on most distortion-based metrics, confirming that classical baselines still provide the highest fidelity in the current setting. However, SIREM consistently improves over its audio-ablated variant, indicating that synchronized speech contributes useful complementary information to reconstruction.

Table~\ref{tab:audio_delta} quantifies this effect: speech conditioning improves all 21 metric--target comparisons, with the largest gains under the gridding target, where audio conditioning yields gains of $+0.048$ SSIM, $+0.96$ PSNR, and $+0.045$ VIF while also reducing LPIPS by $0.018$ and FID by $6.47$.

These results establish an initial benchmark for speech-informed rtMRI reconstruction, showing that synchronized speech serves as a useful auxiliary prior despite not yet matching classical baselines in fidelity.

\begin{table}[!htpb]
\centering
\caption{
Effect of synchronized speech conditioning, measured as the difference between SIREM and SIREM ($w/o$ audio). Positive $\Delta$ is better for SSIM, PSNR, and VIF; negative $\Delta$ is better for HFEN, NRMSE, LPIPS, and FID. Audio conditioning improves all 21 metric--target comparisons.
}
\setlength{\tabcolsep}{12pt}
\label{tab:audio_delta}
\resizebox{\textwidth}{!}{
\begin{tabular}{lrrrrrrr}
\toprule
\textbf{Target} &
$\Delta$SSIM &
$\Delta$PSNR &
$\Delta$HFEN &
$\Delta$NRMSE &
$\Delta$LPIPS &
$\Delta$VIF &
$\Delta$FID \\
\midrule
TV        & +0.018 & +0.37 & -0.035 & -0.004 & -0.008 & +0.019 & -2.69 \\
Gridding  & +0.048 & +0.96 & -0.076 & -0.010 & -0.018 & +0.045 & -6.47 \\
Wavelet   & +0.022 & +0.42 & -0.036 & -0.005 & -0.009 & +0.021 & -3.08 \\
\bottomrule
\end{tabular}}
\end{table}

\begin{figure}[!htpb]
\centering
\includegraphics[width=0.8\linewidth]{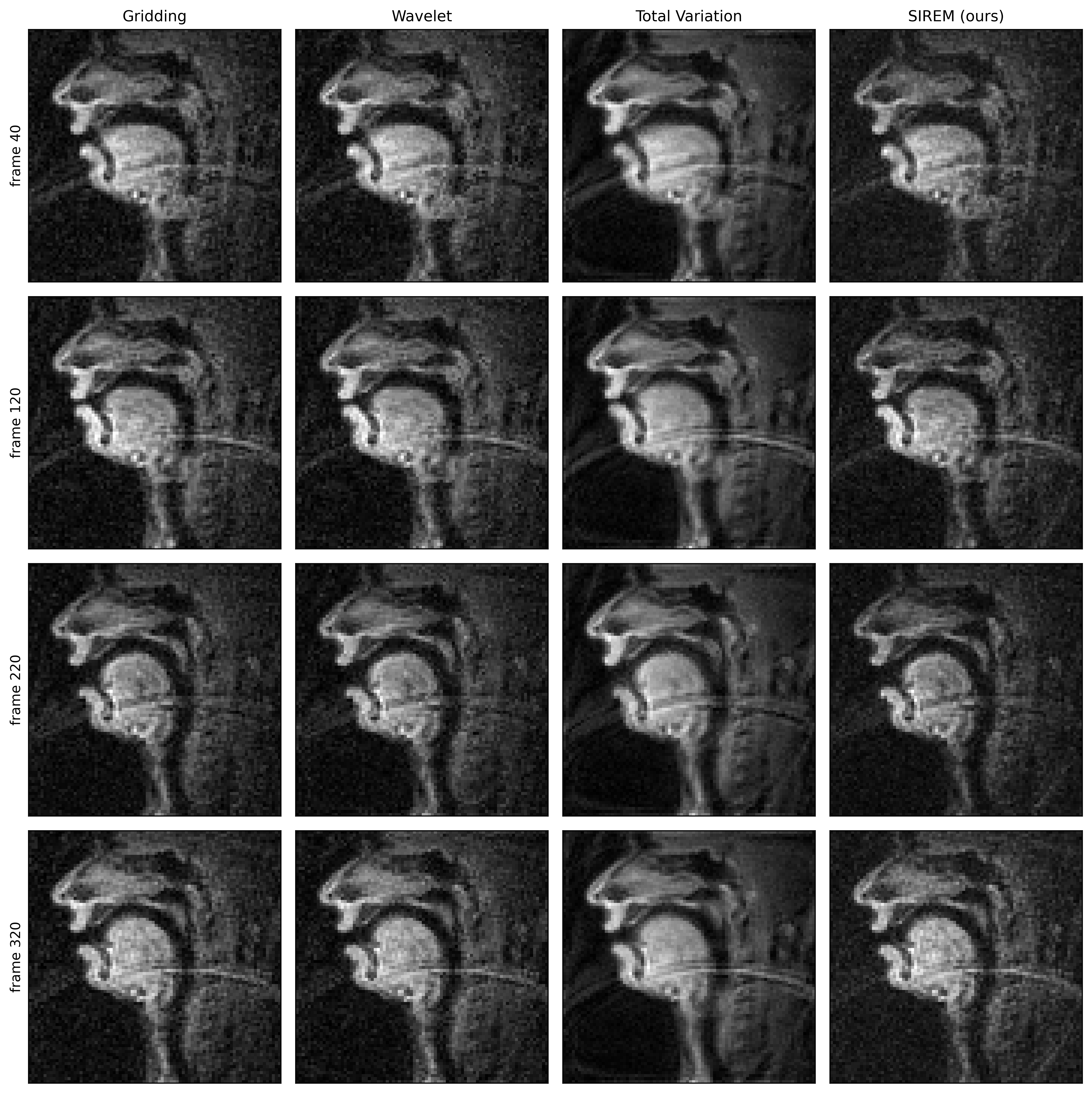}
\caption{Qualitative comparison on five frames from the \emph{Grandfather} passage of a held-out test subject. Gridding shows streak artifacts, Wavelet is smoother, and Total Variation suppresses fine detail; SIREM preserves some articulatory boundaries while operating with feed-forward inference.}
\label{fig:qualitative_comparison}
\end{figure}

Figure~\ref{fig:qualitative_comparison} presents a qualitative comparison of reconstruction methods on representative frames from the \textit{Grandfather} passage of a held-out test subject. Gridding reconstructions exhibit visible streaking and aliasing artifacts caused by undersampling, while wavelet produces smoother reconstructions with reduced artifacts. Total-variation reconstruction further suppresses noise and streaking but also removes some fine anatomical detail, leading to oversmoothed articulatory boundaries. In comparison, SIREM produces reconstructions with sharper tongue and vocal-tract contours than gridding while maintaining a direct feed-forward inference pipeline.

\subsection{Efficiency analysis}

Beyond reconstruction fidelity, an important objective of the proposed method is computational efficiency. Classical baselines such as wavelet reconstruction and total-variation reconstruction rely on iterative optimization, which increases runtime and limits throughput in accelerated settings. In contrast, SIREM performs direct speech-informed reconstruction once the model is trained, avoiding iterative inference at test time. This makes it better aligned with real-time applications, where reconstruction speed is a practical constraint in addition to image quality.

\begin{figure}[!htpb]
    \centering
    \includegraphics[width=0.85\linewidth]{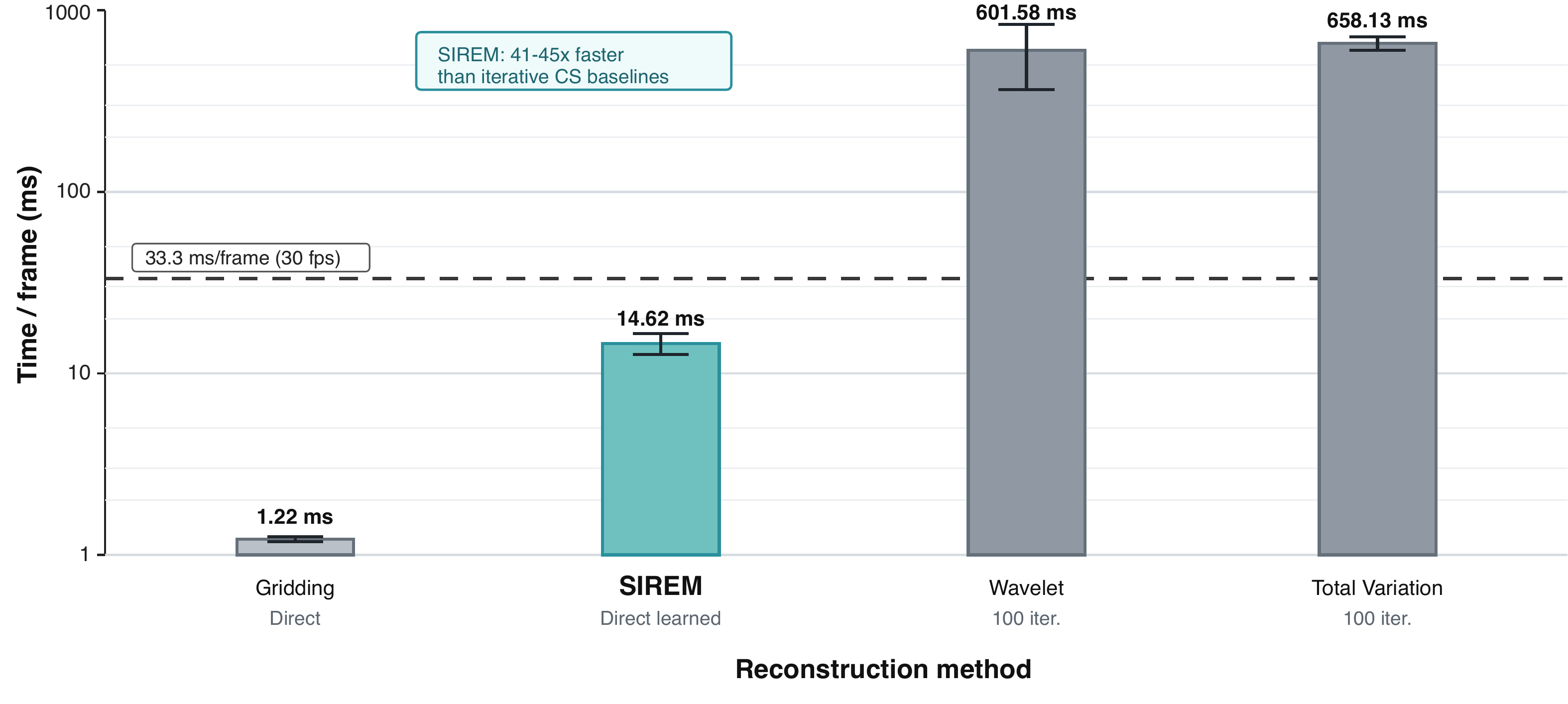}
    \caption{Runtime analysis of reconstruction methods on the test set. Bars show mean time per frame and error bars show sequence-level variation. The dashed line indicates the 30 fps real-time threshold (33.3 ms/frame). SIREM is the only nontrivial reconstruction method evaluated here that operates within the real-time regime, highlighting its efficiency advantage over iterative baselines.}
    \label{fig:efficiency_comparison}
\end{figure}

Figure~\ref{fig:efficiency_comparison} reports throughput, per-frame reconstruction time, iteration count, and inference type averaged over the 63 test sequences. SIREM reconstructs each frame in $\sim$14.6~ms, which is $\sim$41$\times$ faster than Wavelet ($\sim$601.6~ms/frame) and $\sim$45$\times$ faster than Total Variation ($\sim$658.1~ms/frame). Both iterative methods require 100 optimization iterations and remain above 600~ms per frame. Direct gridding is fastest at $\sim$1.2~ms per frame, but it is best interpreted as a lower-bound runtime because it performs a single adjoint pass and produces an aliased reconstruction. Among the nontrivial reconstruction methods evaluated here, SIREM is the only one that operates in a real-time throughput range.
The gridding method (direct NUFFT+SENSE) is the fastest at $\sim$1.2~ms per frame, but this is a trivial lower bound rather than a competitor: gridding solves no inverse problem; rather, it performs a single adjoint pass and produces a aliased reconstruction. Among methods that actually reconstruct, SIREM (ours) is the only one that operates in the real-time range.

\section{Discussion and Conclusions}

This work introduces SIREM, a speech-informed reconstruction framework for rtMRI that uses synchronized audio as an auxiliary prior during reconstruction. Classical baselines remain stronger on most distortion-based metrics, but the results establish an initial benchmark for multimodal speech-informed rtMRI reconstruction and show that direct audio-informed reconstruction is feasible in a substantially higher-throughput regime than iterative baselines. The main contribution of SIREM is therefore not a uniform fidelity gain, but a multimodal reconstruction formulation with a distinct fidelity-efficiency trade-off.

Several limitations remain. The current model uses a fixed segmentation-derived explained-by-audio map rather than a learned fusion predictor, and the spiral-arm profile is evaluated as retrospective soft reweighting rather than as a prospective acquisition policy. The evaluation is also limited to a relatively small subject-independent test set from a single benchmark. Future work should study learned fusion maps, stronger MRI reconstruction backbones, prospective sampling strategies, and targeted ablations of the audio branch, fusion mechanism, and arm profile.

Potential clinical use also requires caution. Faster speech rtMRI reconstruction could reduce scan time and improve practicality, but learned reconstructions should not be used for clinical decision-making without prospective validation across speakers, disorders, scanners, and acquisition settings.

\bibliographystyle{plainnat}
\bibliography{sample}

\appendix

\section{Additional Motivation and Technical Details}

This supplementary material expands the motivation, formulation, and implementation details of the proposed speech-informed reconstruction framework. In particular, it clarifies why synchronized speech constitutes a meaningful prior for vocal-tract MRI reconstruction, provides additional interpretation of the fusion and sampling objectives, and summarizes implementation details omitted from the main paper for space reasons.

\section{Why Speech-Informed Reconstruction is Well Motivated}

The proposed method is based on the observation that speech rtMRI is intrinsically multimodal: the audio waveform and the MRI measurements are recorded simultaneously and arise from the same articulatory process. The acoustic signal is therefore not a side channel, but a consequence of the underlying vocal-tract configuration. Positions of the tongue, lips, jaw, and velum are known to shape the spectral envelope, formant trajectories, and voicing structure of speech. This makes audio informative about at least part of the anatomical configuration visible in rtMRI.

At the same time, the relation between acoustics and anatomy is not uniform across the image. Dynamic articulators such as the tongue and lips are strongly coupled to the produced sound, whereas static anatomy, background tissue, and scanner-related intensity structure are less directly predictable from audio alone. This motivates a spatially varying multimodal model rather than a purely audio-driven one. In other words, synchronized speech should be treated as a structured prior on acoustically informative regions, not as a substitute for the entire reconstruction problem.

This view naturally suggests a decomposition into two complementary components: an audio-driven branch that predicts image content that is largely constrained by acoustics, and an MRI-driven branch that reconstructs image content that must remain anchored to measured k-space data. The explained-by-audio map then determines how strongly each location should depend on either source.

\section{Interpretation of the Fusion Formulation}

Let $x_t$ denote the target frame at time $t$. In SIREM, the audio branch produces the estimate
\begin{equation}
x_t^{a} = F(a_t),
\end{equation}
while the MRI branch produces the estimate
\begin{equation}
x_t^{m} = \mathcal{F}^{-1}(p \odot k_t),
\end{equation}
where $F$ maps synchronized speech to image space, $p$ is the learnable soft spiral-arm profile, and $\mathcal{F}^{-1}$ denotes the adjoint-NUFFT-based reconstruction and coil-combination operator.

Neither branch alone is sufficient. The MRI-only branch ignores the structured multimodal prior contained in the waveform, while the audio-only branch cannot recover image content that is weakly constrained by acoustics. We therefore introduce the spatially varying fusion model
\begin{equation}
\hat{x}_t
=
w_{\mathrm{EbA}} \odot x_t^{a}
+
\left(1-w_{\mathrm{EbA}}\right)\odot x_t^{m},
\label{eq:supp_fusion}
\end{equation}
where $w_{\mathrm{EbA}} \in [0,1]^{H \times W}$ is the explained-by-audio map. Equation~\eqref{eq:supp_fusion} can be interpreted as a convex spatial interpolation between an acoustically informed estimate and a measurement-driven reconstruction. Regions with large values of $w_{\mathrm{EbA}}$ are assumed to be well explained by speech, whereas regions with small values are reconstructed primarily from MRI data.

This construction imposes a useful inductive bias. Rather than forcing the model to decide globally whether a frame should be reconstructed from audio or MRI, it permits local specialization. This is particularly appropriate in speech MRI, where different anatomical regions contribute differently to the acoustic signal.

\section{Basic Property of the Fusion Operator}

The fusion formulation in Eq.~\eqref{eq:supp_fusion} satisfies a simple but useful boundedness property.

\paragraph{Proposition 1.}
Assume $w_{\mathrm{EbA}} \in [0,1]^{H \times W}$ and suppose both $x_t^{a}$ and $x_t^{m}$ take values in $[0,1]^{H \times W}$. Then the fused reconstruction $\hat{x}_t$ defined in Eq.~\eqref{eq:supp_fusion} also lies in $[0,1]^{H \times W}$.

\paragraph{Proof.}
Consider any pixel location $(i,j)$. Since $w_{\mathrm{EbA}}(i,j)\in[0,1]$, we can write
\begin{equation}
\hat{x}_t(i,j)
=
w_{\mathrm{EbA}}(i,j)\,x_t^{a}(i,j)
+
\left(1-w_{\mathrm{EbA}}(i,j)\right)x_t^{m}(i,j).
\end{equation}
Because both branch outputs lie in $[0,1]$, this expression is a convex combination of two values in $[0,1]$. Therefore $\hat{x}_t(i,j)\in[0,1]$. Since this holds for every pixel, $\hat{x}_t \in [0,1]^{H \times W}$. \hfill $\square$

This does not amount to a full theoretical analysis of the method, but it makes explicit that the fusion operator preserves normalized image ranges when both branch outputs are themselves normalized.

\section{Interpretation of the Learned Sampling Objective}

The spiral-arm profile is parameterized by trainable logits $\ell \in \mathbb{R}^{13}$:
\begin{equation}
p = \sigma(\ell),
\end{equation}
where $\sigma(\cdot)$ is the sigmoid function. This yields a differentiable soft weighting of the 13 spiral arms. In the present implementation, these weights do not alter the physical acquisition, but instead define a differentiable reweighting of the available k-space measurements during reconstruction.

The sampling-related regularizers in the training objective have complementary roles. The budget term
\begin{equation}
\mathcal{L}_{\mathrm{budget}}
=
\left(\sum_i p_i - K\right)^2 + \mathrm{mean}(p)
\end{equation}
encourages the effective sampling budget to concentrate around a target number $K$ of active arms. The first term controls the total mass of the soft sampling vector, while the second discourages diffuse, uniformly nonzero weights.

The regularizer
\begin{equation}
\mathcal{L}_{\mathrm{psf}}
=
\mathrm{mean}\left(|\mathcal{F}^{-1}(p)|^2\right)
\end{equation}
penalizes the energy of the inverse Fourier transform of the sampling profile. Intuitively, this discourages pathological arm configurations that would induce undesirable point-spread behavior in the image domain. While this term is heuristic rather than derived from a full acquisition-theoretic analysis, it biases the learned policy toward smoother and more stable weighting patterns.

Finally, the regularizer
\begin{equation}
\mathcal{L}_{\mathrm{mask}}
=
\mathrm{mean}\left(|\mathcal{F}(1-w_{\mathrm{EbA}})|^2\right)
\end{equation}
promotes spatial smoothness in the complement of the explained-by-audio map. Since $w_{\mathrm{EbA}}$ determines where the model should rely on audio versus MRI measurements, this term discourages implausibly noisy or fragmented spatial switching.

\section{Why a Fixed Weighting Map is a Reasonable Starting Point}

In the main experiments, the explained-by-audio map $w_{\mathrm{EbA}}$ is not learned jointly but is provided by segmentation-derived masks. This choice was made for two reasons. First, the USC Annot-16 subset is small, and learning both the image prediction branch and the fusion map jointly would introduce additional ambiguity in an already low-data regime. Second, the fixed masks impose an interpretable prior: regions associated with dynamic articulators are explicitly marked as more likely to be predictable from acoustics.

This design should be understood as a first instantiation of the multimodal reconstruction idea rather than as its final form. A learned weighting map may ultimately be preferable, but the fixed-mask setting is a useful starting point because it isolates the multimodal fusion mechanism from the challenge of simultaneously discovering where that fusion should occur.

\section{Additional Dataset Details}

The experiments are based on the USC Annot-16 subset of the USC 75-speaker speech rtMRI corpus~\cite{lim2021multispeaker}. USC-16 contains 16 speakers with manual articulatory annotations and serves as a benchmark for segmentation-based analyses. We use a subject-independent split with 10 training subjects, 2 validation subjects, and 4 test subjects. The main paper reports aggregate results across the held-out test set.

The full USC 75-speaker dataset comprises midsagittal rtMRI acquisitions of speech production collected on a 1.5T scanner with synchronized audio, raw k-space data, and reconstructed 2D image sequences. The data include a wide range of linguistically motivated speech tasks and capture the full moving vocal tract at high temporal resolution.

\section{Additional Preprocessing Details}

Reference reconstructions and segmentation masks are aligned to each 13-arm acquisition window by selecting the temporally centered frame within the corresponding interval. This is necessary because the reference sequences and masks are available at the 2-arm rate, whereas the reconstruction pipeline operates on 13-arm windows. The associated waveform segment is extracted around the center of the same interval using a symmetric context window.

All image-domain inputs are resized to $84 \times 84$ and normalized to $[0,1]$. K-space values are normalized by their maximum magnitude before frame formation, and trajectories are scaled to the reconstruction grid. Coil sensitivity maps are estimated once per utterance and reused across frames of that utterance.

\section{Additional Implementation Details}

The audio-driven branch uses a frozen HuBERT encoder (\texttt{facebook/hubert-base-ls960}) followed by a lightweight multilayer decoder. Only the decoder and the sampling logits are optimized in the reported experiments. Optimization is performed with AdamW using a learning rate of $10^{-4}$, weight decay of $10^{-5}$, batch size 8, cosine annealing over 20 epochs, and gradient clipping with maximum norm 1.0. Model selection is based on validation PSNR.

\end{document}